\documentclass[12pt]{article}

\usepackage{epsfig,amsmath,float,fullpage}
\usepackage{wrapfig}
\usepackage{url}
\usepackage{sidecap}
\usepackage{multirow}
\usepackage{amsmath}
\usepackage{graphicx}
\usepackage{amsthm}
\usepackage{amssymb}
\usepackage{epstopdf}
\usepackage{setspace}

\newcommand{\ds}{\displaystyle}
\newcommand{\rmp}{R_{1,0}}
\newcommand{\mrm}{\mathrm}
\newcommand{\xbza}{X_i'\vec\beta + Z_{ij}'\vec\alpha}
\newcommand{\xbzal}{X_i'\vec\beta + Z_{i\ell}'\vec\alpha}
\newcommand{\xb}{X_i'\vec\beta}
\newcommand{\za}{Z_{ij}'\vec\alpha}

\newcommand{\delgam}[1]{\Delta_{i#1}+\Gamma_{i#1}}
\begin{document}
\title{Models with time-varying predictors for meningitis in Navrongo, Ghana}

\renewcommand{\arraystretch}{0.65}
\author{\small Yolanda Hagar, Mary Hayden, Abudulai Adams Forgor, Patricia Akweongo, \\
\small Abraham Hodgson, Christine Wiedinmyer, Vanja Dukic
\footnote{
Yolanda Hagar  is a postdoctoral researcher in applied mathematics at the
University of Colorado at Boulder. Mary Hayden and Christine Wiedinmyer are scientists at the National Center of Atmospheric Research in Boulder, Colorado. Abudulai Adams Forgor is the medical superintendent at the War Memorial Hospital, Ghana Health Service in Navrongo, Ghana.  Patricia Akweongo is a senior lecturer at the Department of Health Policy, Planning and Management, School of Public Health at the University of Ghana in Accra, Ghana.  Abraham Hodgson is director of the Research and Development Division of Ghana Health Services, Accra, Ghana. Vanja Dukic is a professor in applied mathematics at the University of Colorado at Boulder.  
\newline
Vanja Dukic (email: {\tt Vanja.Dukic@colorado.edu}) 
}}

\date{  }
\maketitle

\thispagestyle{empty}
\baselineskip 12pt

\begin{abstract}
\noindent
The ``meningitis belt" is a region in sub-Saharan Africa where annual outbreaks of meningitis occur, with large epidemics observed cyclically.  While we know that meningitis is heavily dependent on seasonal trends (in particular, weather), the exact pathways for contracting  the disease are not fully understood and warrant further investigation.  This manuscript examines meningitis trends in the context of survival analysis, quantifying underlying seasonal patterns in meningitis rates through the hazard rate for the population of Navrongo, Ghana. We compare three candidate models: the commonly used Poisson generalized linear model, the Bayesian multi-resolution hazard model, and the Poisson generalized additive model. We compare the accuracy and robustness of the models through the bias, RMSE, and the standard deviation.  We provide a detailed case study of meningitis patterns for data collected in Navrongo, Ghana.
\\

\noindent
{\em Key Words:} survival analysis, hazard rate, count data, multi-resolution hazard, changing at-risk population, time-varying covariates
\end{abstract}

\newpage
\section{Introduction}

Within the ``meningitis belt", a region in sub-Saharan Africa stretching from Ethiopia to Senegal (also known as the Sahel), outbreaks of meningococcal meningitis happen annually during the dry season, with large epidemics occurring every 2 to 10 years \cite{lapey, greenwood99}.  The dynamics of transmission of meningitis in the Sahel are not fully explained, as the causes are likely due to a variety of factors, including socio-economic status, migration, environmental circumstances, and the introduction of new strains of the disease.  While meningitis can be controlled through vaccination, this strategy is dependent on the identification of populations that are at risk and periods of time when transmission is most likely.  Meningitis is a serious illness (with approximately 1 in 10 cases being fatal \cite{moore89}), however, the actual number of diagnosed meningitis cases can be small when compared to the number of people at risk (ranging from 10 to 1000 infections per 100,000 people \cite{greenwood87}), making inference about the wide range of contributing factors difficult.  

Meningitis is highly dependent on seasonal trends, with larger epidemics occurring cyclically over multiple year time-spans.  Several studies support the idea that weather conditions greatly impact the transmission of meningitis, with higher incidence rates occurring between December and May when the weather is dry \cite{abdussalam, besancenot, cheesbrough, cuevas, greenwood99, greenwood84,  lapey, molesworth02, moore92, molesworth03, roberts, sultan05a, sultan05b, sultan07, thomson, yaka}.  However, even with reliable weather forecasting, patterns in yearly and cyclic meningitis rates cannot be entirely explained by changes in climate, as there are many other temporal factors that co-vary with weather patterns (such as changes in social activities, patterns of migration, and variations in cooking methods).  These other trends can be more difficult to measure but may have an equally significant impact on the transmission of meningitis.  Therefore, statistical models quantifying trends in meningitis outbreaks must be able to account for seasonal covariates as well as yearly trends in the disease due to latent variables.

Many statistical approaches have been used to quantify meningitis rates in other areas of the meningitis belt.  \cite{beresniak} created a Bayesian network to model how meningitis incidence rates change based on dependent relationships among different districts in Niger.  This model is a graphical tool that showed the probabilistic relationships of contracting meningitis among different regions.  \cite{ghanaCycle} used a wavelet method to compare and contrast different time series of counts for various countries in the meningitis belt, examining both yearly trends and cyclical epidemics. Neither of these two analyses incorporated weather covariates or other predictors in the modeling of meningitis counts, but they did characterize general trends and peaks in the disease over many decades.  Other types of modeling used a generalized linear model approach, assuming a negative binomial distribution for meningitis counts \cite{CPGarcia2}.  Work by \cite{Stanton} compared different types of models, including dynamic regression models (using a Bayesian framework), and a three-state Markov chain model estimating the probability of changing between different incidence risk levels.  In \cite{diggle2010}, the authors examined individual and group-level data, and modeled the risks based on both individual outcomes and spatially related measures using a Poisson cluster point process.  In a similar vein, \cite{agier} discretized weekly incidence rates into latent states based on different epidemiological thresholds, modeling the probability of transitioning between states using a multi-state Markov model, accounting for climate variables and dependencies based on a variety of factors.  These methods are useful, and account for both changes in climate and spatial information over time.  However, the data for the analyses mentioned above mostly come from Niger, which experiences large epidemics every 8 to 12 years and generally has a higher incidence rate than Ghana, which has not had a cyclic epidemic pattern since the 1960's \cite{ghanaCycle}.  In this manuscript, we examine how three similar models behave under circumstances with low event counts and sparse amounts of information, as is the case with the meningitis counts for our data from Navrongo, Ghana and the surrounding Kassena-Nankana District.

\subsubsection*{Meningitis Data}
The data we examine in this manuscript are 364 meningitis cases, aggregated as monthly counts over an 11-year period from 1998 to 2008 in Navrongo and the surrounding Kassena-Nankana (K-N) District in Northern Ghana.  Weather measurements are also reported, including daily dust status, number of sunshine hours, maximum and minimum daily temperature, relative humidity, rain quantity, and wind speed.  Because the meningitis counts are monthly aggregated figures, the weather variables are provided as monthly averages or percentages.  (See \cite{dataref} and \cite{jabes} for details on the data set.) 

This particular data set from Navrongo was analyzed by \cite{jabes}, using a generalized additive model (GAM; \cite{GAM}) assuming a Poisson distribution on the meningitis counts, allowing for a smooth function to account for underlying baseline trends in the seasonality of meningitis while estimating the impact of changes in weather.  The authors explore optimal models for describing patterns in meningitis by using biological rationale, reduction of collinearity among variables, and the Aikaike Information criteria (AIC; \cite{refAIC}).   

We expand on this work, and present our examination of this data in the context of survival analysis, quantifying underlying seasonal patterns in meningitis rates through the hazard rate while estimating the impact of the time-varying effects of weather. Our goal is to compare differences in estimation among three suitable models: 1) The ubiquitous Poisson generalized linear model (GLM;\cite{GLM}), also referred to in some literature as the log-linear model; 2) The Bayesian multi-resolution hazard model (MRH; \cite{Bouman2, Bouman, Yprune, Dignam, Dukic, BCRT}), which has a likelihood function identical to the first model, but uses a tree-like prior that allows for sharing of information across time periods, making it an ideal method for estimation when the number of observed failures are low; and 3) The Poisson generalized additive model (GAM; \cite{GAM}), which is similar to the previous two models, but the hazard rate is estimated with a smooth function instead of a piecewise-constant function. We employ the well-established link between the log-linear model and the piecewise exponential hazard model (\cite{Holford80, Laird}), as time-varying covariates and a changing (i.e. not monotonically decreasing) at-risk population are easily incorporated into the piecewise exponential model.  

This paper is organized as follows: In Section 2, we compare theoretical differences among the three different models (MRH, GLM, and GAM) and examine how they relate to the piecewise exponential model, and in Section 3 we examine the variance for the Poisson GLM and MRH model.  In Section 4, we compare the robustness and accuracy of the three models through simulations, and in Section 5, we examine the Navrongo data set.  We conclude in Section 6 with a discussion. 

\section{Hazard Models for Meningitis}

Hazard modeling is commonly associated with survival analysis.  However, the hazard rate function, defined as $h(t) = \lim_{\Delta \to 0} {P(t \le T < t+\Delta \mid T \ge t)}/{\Delta} = {f(t)}/{S(t)}$ (where $S(t)$ is the survival function and $f(t) = -S'(t)$), is also a convenient method for characterizing a process over time.  In this manuscript, the hazard rate is used to describe counts of events (such as the number of meningitis cases) over the course of a year for an inhomogeneous Poisson process.  While there are a plethora of methods for estimation of the hazard rate (for examples, see \cite{Ibrahim, kleinmoesch,KalbPrenticeBook}), a simple and convenient method for describing the time-varying pattern is with a piecewise constant function.  In general theory, this is known as the piecewise exponential (PE) hazards model \cite{Glasser,Prentice73,Holford76,Holford80,Laird}, which easily accommodates time-varying covariates (such as monthly weather measurements) as well as a changing number of subjects at risk (such as the increase in the population of Navrongo between 1998 and 2008).  We are interested in examining how this model (represented through the Poisson generalized linear model \cite{GLM}) compares to the modern, Bayesian MRH model, which can also be expressed as a PE model but has a prior structure that lends itself to estimation in the instance of small observed failures.  We contrast these two models to a Poisson GAM, as it estimates the hazard rate with a smooth function while still accommodating time-varying weather covariates and a changing at-risk population.

In the following subsections, we compare the likelihood formulation of the Poisson GLM, the MRH model, and the Poisson GAM.  All models have an underlying function that accounts for seasonal baseline meningitis risk, accounting for trends that cannot be fully explained by weather covariates.  In addition, all three models can accommodate time-varying weather variables, as well as a changing population at risk.  

\subsection*{Piecewise Constant Hazard Models}

Below, we review and define notation for PE models and then contrast them to our models in the following sections.  As we will show, both the Poisson GAM and the MRH model have likelihoods identical to the piecewise exponential model, with a similar form for the Poisson GAM.  

In the context of survival analysis, a ``failure" is defined as the contraction of meningitis, and the ``survival time" represents the time from the beginning of the seasonal year (i.e. June) until contraction of meningitis.  Subjects are censored due to death or because the 11-year observation period is over.  Let $T_i$ represent the survival time of the $i^{th}$ individual, and let $\delta_i$ be the censoring indicator, which equals `1' if the failure is observed and `0' otherwise.  For a piecewise approximation of the hazard rate, the study period $[t_0, t_J]$ is divided into $J$ intervals, $(t_{j-1}, t_j),  j = 1, \dots, J,$ and a constant baseline hazard rate is estimated within each time interval (or ``bin").   Let $\tau_{ij}$ denote the time subject $i$ spent in interval $j$, such that 
\begin{equation}
	\tau_{ij} = \begin{cases}
		0&\text{ if }T_i < t_{j-1}\\
		T_i-t_{j-1}&\text{ if } T_i \in [t_{j-1}, t_j]\\
		t_j-t_{j-1}&\text{ if } T_i >  t_j.
		\end{cases}
\end{equation}
Similarly, we can define $\delta_{ij}$ such that it equals `1' if individual $i$ had an observed failure in bin $j$, and is `0' otherwise, noting that $\delta_i = \sum_{j=1}^J \delta_{ij}.$  The baseline cumulative hazard function is defined as $H_0(T_i) = \sum_{j = 1}^J \tau_{ij}\lambda_j,$ where $\lambda_j$ is the constant baseline hazard rate in bin $j$.  Covariates (both time-varying and time-independent) can easily be incorporated into the model by defining the cumulative hazard as
\begin{eqnarray*}
	H(T_i \mid X_i, Z_i, \vec\beta, \vec\alpha, \vec\lambda) = \left(\sum_{j = 1}^J \tau_{ij}\lambda_j\exp\{\za\}\right)\exp\{\xb\},
\end{eqnarray*} 
where for subject $i$, $X_i$ is a $1\times p$ vector of $p$ time-independent (e.g. as gender) covariate measures, $Z_{ij}$ is a $1\times q$ vector of $q$ time-varying (e.g.  monthly temperature) covariate measures in bin $j$, and $\vec\beta$ and $\vec\alpha$ are, respectively, the effects of the covariates on the time to survival.  (Note that this formulation assumes that the time-varying covariates change only at the boundaries of each bin, which is a valid assumption for our data but may require more advanced techniques with other data sets.)  
\subsubsection*{Likelihood Function}

Using the notation above, the log-likelihood function for subject $i$ can be written as:
\begin{align*}
	\log(L_i  \mid T_i, \delta_i, X_i, Z_i, \vec\beta, \vec\alpha, \vec\lambda) & =\delta_i\log h(T_i \mid X_i, Z_i, \vec\beta, \vec\alpha, \vec\lambda)-H(T_i \mid X_i, Z_i, \vec\beta, \vec\alpha, \vec\lambda)\\
		& = \sum_{j=1}^J\left\{\delta_{ij}\left(\log\left(\tau_{ij}\lambda_{j}\right)+\xbza\right) - \tau_{ij}\lambda_{j}\exp\{\xbza\}\right\}.
\end{align*}
The standard piecewise exponential model can be easily modified to accommodate grouped observations (i.e. multiple failures at the same time point) as well as a changing population of subjects at risk over time.  In this instance, $i$ denotes a subject \textit{group}, such as the subjects who contract meningitis within a year $i$.  We define $\Delta_{ij}$ as the number of subjects in group $i$ who failed in bin $j$ and $\Gamma_{ij}$ as the number of subjects right-censored in bin $j$.  In this instance, the log-likelihood contribution for subjects from group $i$ who fail in bin $j$ (either as an observed or censored failures) can be rewritten as: 

\begin{align}
	& \log(L_{ij}  \mid T_i, X_i, \mathbf{Z}_i, \vec\beta, \vec\alpha, \vec\lambda) \nonumber\\
	& \qquad = \Delta_{ij}\left(\log\left(\tau_{ij}\lambda_{j}\right)+\xbza\right) - \left(\delgam{j}\right)\sum_{\ell=1}^j\tau_{i\ell}\lambda_{\ell}\exp\{\xbzal\}\label{eqn:grpLLsec},
\end{align}
noting that $\tau_{i\ell}$ is equal to zero for all bins past the failure bin $j$.  While piecewise exponential models do not require that all time bins be the same width, in our analysis we divide the study time into equal lengths since our data is reported monthly.  If the bin width is denoted as  $\omega$ (i.e. $\omega = t_j-t_{j-1}, j = 1, \dots, J$), the cumulative hazard in equation (\ref{eqn:grpLLsec}) for all subjects at risk in group $i$ across all $J$ bins can be rewritten as:
\begin{align}\label{eqn:cumulfinal}
	\sum_{j = 1}^J & \left(\delgam{j}\right)\sum_{\ell=1}^j\tau_{i\ell}\lambda_{\ell}\exp\{\xbzal\} \nonumber\\
		& = \exp\{\xb\}\sum_{j=1}^J\omega\lambda_{j}\exp\{\za\}\left(\frac{\tau_{ij}}{\omega}\left(\delgam{j}\right) + \sum_{\ell = j+1}^J\left(\delgam{\ell}\right)\right)\nonumber\\
		& = \exp\{\xb\}\sum_{j=1}^J\omega\lambda_{j}\exp\{\za\}\left(\frac{\tau_{ij}}{\omega}\left(\delgam{j}\right) + \sum_{\ell = j}^J\left(\delgam{\ell}\right)-\left(\delgam{j}\right)\right)\nonumber\\
		& = \exp\{\xb\}\sum_{j=1}^J\omega\lambda_{j}\exp\{\za\}\Phi_{ij},
\end{align}
with $\Phi_{ij} =  \frac{\tau_{ij}}{\omega}\left(\delgam{j}\right) + N_{ij}-\left(\delgam{j}\right)$.  Note that $N_{ij} = \sum_{\ell = j}^J\left(\delgam{\ell}\right)$ represents the number of subjects at risk in group $i$ at the beginning of bin $j$, and that this formulation of the model allows for new subjects to become at-risk over the course of the observation period.  

The log-likelihood for group $i$ can therefore be written as:
\begin{align}\label{eqn:LLfinal}
	& \log(L_{i}  \mid T_i, X_i, \mathbf{Z}_i, \vec\beta, \vec\alpha, \vec\lambda) = \sum_{j = 1}^J\Delta_{ij}\left(\log\left(\tau_{ij}\lambda_{j}\right)+\xbza\right)	
	- \exp\{\xb\}\sum_{j=1}^J\omega\lambda_{j}\exp\{\za\}\Phi_{ij}.
\end{align}
In the next sections, we will show that the Poisson GLM and the MRH model have likelihood functions equivalent to the one in equation (\ref{eqn:LLfinal}), and that the Poisson GAM is similar, with a smooth approximation for the baseline hazard rate.

\subsection{Poisson GLM model} \label{sec:poissglm}
As shown in \cite{Holford80} and \cite{Laird}, the piecewise exponential model is equivalent to a Poisson regression model, which assumes the number of subjects who contract meningitis within each bin follows a Poisson distribution.  Use of the Poisson model in a survival analysis context assumes that the number of failures within each bin of time is independent.  While this assumption cannot possibly hold in the standard survival analysis setting (as failure of a subject in an early bin is directly related to the number of subjects that can fail in later bins), in this particular context the at-risk population is constantly changing, making the assumption of independence between bins more viable.  In addition, as pointed out in \cite{Laird} and \cite{Holford80}, this method can be used to derive an equivalent log-likelihood function, as we do below.  

\subsubsection*{Likelihood function}
Let $y_{ij}$ represent the number of observed failures from group $i$ in bin $j$. Then, $y_{ij}$ comes from a Poisson distribution with mean $\mu_{ij}$, such that
$$\mu_{ij} = W_{ij}\lambda_{j}\exp\{\xbza\},$$
where $W_{ij}$ is the total time spent by all subjects from group $i$ in bin $j$, and can be re-expressed as 
\begin{align*}
	W_{ij} & = \left(\delgam{j}\right)\tau_{ij} + \left(N_{ij} - \left(\delgam{j}\right)\right)\omega = \omega \Phi_{ij}.
\end{align*}
In the expression above,  $\left(\delgam{j}\right)\tau_{ij}$ represents the length of time the subjects who failed in bin $j$ (censored or not) spent in bin $j,$ and $\left(N_{ij} - \left(\delgam{j}\right)\right)\omega$ represents the amount of time subjects who survived beyond bin $j$ spent in bin $j$ (with $\tau_{ij} < \omega$ for the subjects who failed in bin $j$).  The log-linear model (i.e. the Poisson GLM) can be expressed as 
\begin{eqnarray}\label{eqn:glmregress}\log \mu_{ij} = \log W_{ij}+\log\lambda_{j} + \xb + \za,\end{eqnarray}
where $\lambda_{j}$ represents the baseline seasonal trends in meningitis rates (i.e. the baseline hazard rate).

In this instance,  the log-likelihood for group $i$ over all $J$ bins written as:
\begin{align}\label{eqn:LLglm}
	\log L_{i,Poiss}(y_i \mid&  X_i, Z_i, \vec\beta, \vec\alpha, \vec\lambda) = \sum_{j = 1}^J \left\{y_{ij}\log\mu_{ij} - \log\left(y_{ij}!\right)-\mu_{ij}\right\}\nonumber\\	
		& \propto \sum_{j=1}^J y_{ij} \left(\log\lambda_{j}+\xbza\right) - \exp\{\xb\}\sum_{j = 1}^J\omega\lambda_{j}\exp\{\za\}\Phi_{ij}
\end{align}
Equation (\ref{eqn:LLglm}) is equivalent to equation (\ref{eqn:grpLLsec}) by noting that the number observed failures in each bin, $y_{ij}$, is equal to $\Delta_{ij}$.  Estimation of the parameters is done through iteratively weighted least squares (IWLS) using standard GLM methods \cite{GLM}, with an offset value equal to $\log(W_{ij})$ for each observation.

\subsection{Multi-resolution Hazard Model} 
The multi-resolution hazard (MRH) model, shown in \cite{Bouman2, Bouman, Dukic, Dignam, BCRT, Yprune} and used in \cite{sadm} is a Bayesian piecewise-constant exponential hazard model that is  based on a tree-like, wavelet-based multi-resolution prior on the hazard function, which allows for scalability and consistency across different time scales (i.e minutes, weeks, years, etc).  The MRH model is closely related to the Polya tree \cite{ferguson74, polyaLavine}. The Polya-tree prior is an infinite, recursive, dyadic partitioning of a measurable space $\Omega$, although in practice this process is terminated at a finite level $M$, resulting in ``finite" Polya trees.  Polya trees have been adapted for modeling survival data in a number of ways (for example, see \cite{polyaSurvMuliere, polyaSurvHansonJohnson, polyaSurvHanson, polyaSurvZhao}), including stratified Polya tree priors (\cite{polyaStratifyZhao}), Polya trees with fused bins (\cite{polyaOPTWong}), and smoothed Polya tree priors \cite{polyaRPTNB}).  The MRH prior is a type of Polya tree; it uses a fixed, pre-specified partition, and controls the hazard level within each bin through a multi-resolution parameterization.  This parameterization allows parameters to differ across bins and levels of the tree in such a way that  the marginal priors at higher levels of the tree are the same, regardless of the priors at lower levels of the tree. 

In the MRH model, the hazard rate is approximated in a piecewise fashion with a set of corresponding hazard increments,  $d_{j}, j = 1,\dots, J$, where each $d_j$ represents the aggregated hazard rate over the $j^{th}$ bin.  (i.e. $d_{j} = \int_{t_{j-1}}^{t_{j}} h(s)ds$ in standard survival notation).  We assume that $J=2^{M}$, where $M>0$ and can be chosen in a variety of ways; for example, using model selection criteria as in \cite{Bouman}, or using clinical input, as in \cite{Dignam}.   We then define ``split parameters" $R_{m,p} = H_{m,2p}/H_{m-1,p}, m = 1,2,\dots,M-1, p = 1,\dots,2^m-1$, where $H_{m,p}$ is recursively defined such that $H_{m,p}\equiv H_{m+1,2p}+H_{m+1,2p+1}$.  These split parameters determine the shape of the hazard rate, and are between 0 and 1.  A Gamma($a, b$) prior is placed on $H$, and the prior for $R_{m,p}$ is $\mathcal{B}e(2\gamma_{m,p}k^{m}a,2(1-\gamma_{m,p})k^{m}a)$. The $d_j$ in the final time resolution are estimated using the split parameters and the estimate of the cumulative hazard.  

\subsubsection*{Likelihood function}

Using the notation of the MRH model, we can write the cumulative hazard function as:
\begin{align*}
	H_{MRH}(T_i \mid X_i, Z_i, \vec \beta, \vec \alpha, H, \vec R_{m,p}) = \exp\{\xb\}\sum_{j = 1}^J \frac{\tau_{ij}}{\omega}d_j\exp\{\za\}.
\end{align*}
For the subjects in group $i$, the log-likelihood function becomes:
\begin{align}
	\log L_{i, MRH}(T_i & \mid X_i, Z_i, \beta, \alpha)= \sum_{j = 1}^J \Delta_{ij}\left(\log d_j + \xb + \za\right) 
			- \exp\{\xb\}\sum_{j=1}^Jd_j\exp\{\za\}\Phi_{ij}\label{eqn:LLmrh}.
\end{align}
The log-likelihood in equation (\ref{eqn:LLmrh}) is proportional to the log-likelihood in equation (\ref{eqn:LLfinal}) (and therefore equivalent to equation (\ref{eqn:LLglm}), with $d_j = \lambda_{j}\omega.$ 

\subsubsection*{Parameter estimation}\label{sec:estMRH}

All parameters in the model are estimated using Markov chain Monte Carlo.  This Bayesian approach to modeling the hazard rate is what provides us with the flexibility to estimate hazard rate while remaining computationally feasible, even when the number of observed failures (in this case, the number of observed meningitis cases) is small.

The algorithm consists successive sampling from the following full conditional distributions:
\begin{enumerate}
\item 
The full conditional for $H$ is  a gamma density with the shape parameter
$a + \sum_{i=1}^n\sum_{j = 1}^J\Delta_{ij}$, and rate parameter $b^{-1} + \sum_{i=1}^n\sum_{j=1}^J \exp\left(\xbza\right)\mathcal{F}_j\Phi_{ij},$
where  \begin{eqnarray}\mathcal{F}_j = d_j/H\label{eqn:Fj}\end{eqnarray} represents the combination of $R_{m,p}$ parameters needed to calculate $d_j$.
	
\item 
The full conditional of each $R_{m,p}$ is:
\begin{equation*}\label{Rpost}
\begin{array}{l}
{ R_{m,p}^{2\gamma_{m,p} k^m  a -1}(1-R_{m,p})^{2(1-\gamma_{m,p} )k^m  a -1} }\prod_{i = 1}^nL_i(T_i \mid X_i, Z_i,\vec\beta, \vec\alpha),
\end{array}
\end{equation*}
where the likelihood function is from equation (\ref{eqn:LLmrh}).

\item 
With a $\mathcal{N}(0,\sigma_{\beta_s}^2)$ prior (with a fixed variance) on  each baseline covariate effect,  $\beta_s \,\, (s = 1, \dots, p$), each has the following full conditional distribution:
 \begin{align*}
\pi( \beta_s | \beta_s^{-}) &\propto  \ds 
\Pi_{i= 1}^n\Pi_{j =1}^J \Big[\mrm{exp}\left\{\Delta_{ij}X_{is}\beta_s\right\}\ds \times \exp\left\{-d_j \exp(\xbza)N_{ij}\right\}\Big]
		\exp\left\{-\frac{\beta_s}{2\sigma_{\beta_s}^2}\right\}
\end{align*}

\item 
Similarly, with a $\mathcal{N}(0,\sigma_{\alpha_s}^2)$ prior (with a fixed variance) on  each time-varying covariate effect,  $\alpha_s \,\, (s = 1, \dots, q$), each has the following full conditional distribution:
 \begin{align*}
\pi( \alpha_s | \alpha_s^{-}) &\propto  \ds 
\Pi_{i= 1}^n\Pi_{j =1}^J \Big[\mrm{exp}\left\{\Delta_{ij}Z_{ijs}\alpha_s\right\}\ds \times \exp\left\{-d_j \exp(\xbza)N_{ij}\right\}\Big]\exp\left\{-\frac{\alpha_s}{2\sigma_{\alpha_s}^2}\right\}
\end{align*}

\item 
The full conditional distributions for the remaining parameters ($a,b,k$ and $\gamma_{m,p}$) are the same as with the classical MRH and can be seen in Appendix A.

\end{enumerate}

\subsection{Poisson Generalized Additive Model} \label{sec:poissgam}

The likelihood functions of the Poisson generalized additive model (GAM) and the piecewise exponential are not identical, but are similar.  The main difference is that in the generalized additive model, a smoothed function $g(t)$ is used to represent the baseline hazard rate.  This model (initially used to describe meningitis patterns in this data set in \cite{jabes}) is similar to the log-linear model shown in equation (\ref{eqn:glmregress}): 
\begin{eqnarray}\label{eqn:gamregress} \log \mu_i(t) = \log(W_i(t)) + g(t) + \xb + Z_i(t)\vec\alpha,\end{eqnarray}
where $g(t)$ is a smooth function that corresponds to the baseline hazard rate, and $Z_i(t)$ represents the value of the time-varying covariate at time $t$ for group $i$. Note that $\lambda_j$ corresponds to the average value of $g(t)$ over $t\in[t_{j-1}, t_j]$, $\lambda_j =\frac{1}{\omega}\int_{t_{j-1}}^{t_j} g(t)dt.$  Similarly, $Z_{ij}$ represents the average of $Z_i(t)$ over $t\in[t_{j-1}, t_j],$ and $W_{ij}$ is the average amount of time spent by the $i^{th}$ group in bin $j$, such that $W_{ij} = \int_{t_{j-1}}^{t_j} W_i(t)dt.$

\section{Impact of MRH Prior on Estimation Efficiency} \label{sec:varcompare}

When the number of observations increases, Bayesian models tend to perform identically to their frequentist counterparts as the effects of the priors diminishes.  Thus, when the observed number of meningitis cases is large (within each interval of time), we expect the Poisson GLM and MRH models to perform identically both in accuracy of parameter estimation as well as efficiency.  However, we are interested in comparing the two models when the number of observed failures is small (particularly when compared to the number of people at risk), as is the case with meningitis rates in Navrongo.  

Of particular interest is the variance for the $J$ piecewise estimates of the hazard rate.  In this section, we compare the variances of the Poisson GLM and MRH models through the negative expected values of the second derivatives of the log-likelihood and the posterior distribution.  For notational simplicity, the calculations are performed on a reduced version of the piecewise model: We assume that there are only two bins (i.e. $J = 2$), that subject groups fail in the middle of the bin (i.e. $\tau_{ij} = 0.5$), and that the length of the bin is equal to 1 (i.e. $\omega = 1$) so that $d_j = \lambda_{j}.$  We omit covariates, assume there is only one subject group (i.e. $n = 1$), and omit censoring mid-study.   However, these results can be extended to more complex circumstances.

In the instance of two bins and one subject group, omitting the group-specific notation, the log-likelihood function for both the Poisson GLM and MRH model becomes
\begin{align*}
	\log L(T \mid \vec{\lambda}) &  =
	\left(\Delta_{1}\log(\lambda_1) - \lambda_1\left(N_1-0.5\Delta_{1}\right)\right)+\left(\Delta_{2}\log(\lambda_2) - \lambda_2\left(N_2-0.5\Delta_{2}\right)\right).
\end{align*}
In this model, $\lambda_1 = d_1 = H\times R_{1,0}$.

Under the Poisson model, the variance for $\hat{\lambda}_1$ and $\hat{\lambda}_2$ can be easily calculated as $\Sigma_{GLM} = \left[-E(H_{GLM})\right]^{-1},$ (where $\Sigma_{GLM}$ is the negative expected value of the second derivative of the log-likelihood function), such that 
$$var(\hat{\lambda}_1) = \frac{\lambda_1^2}{N_1(1-\exp\{-\lambda_1\})},\text{       } var(\hat{\lambda}_2) = \frac{\lambda_2^2}{N_1(1-\exp\{-\lambda_2\})}.$$  Note that  $E\Delta_j = R_jP(\text{fail in bin $j$}) = R_j\left(1-\exp(-d_j)\right).$  The 
Similarly, the negative expected value of the second derivative of the posterior for the MRH model is equal to: 
\begin{align*}
		-E(H_{MRH}) & =\left[ \begin{array}{cc}
			\frac{N_1\left(1-e^{-d_1}\right)+N_2\left(1-e^{-d_2}\right)+a-1}{H^2}&\phi\\
		\phi&\frac{N_1\left(1-e^{-d_1}\right)+ak}{R_{1,0}^2}+\frac{N_2\left(1-e^{-d_2}\right)+ak}{(1-R_{1,0})^2}\\
	\end{array}\right],\\
\end{align*}
where $\phi = N_1\left(1- 0.5(1-e^{-d_1})\right) - N_2\left(1-0.5(1-e^{-d_2})\right).$  Using the Laplace approximation \cite{Laplace}, the variance of $\hat{d}_1$ and $\hat{d}_2$ (the means of the marginal posterior distributions for $d_1$ and $d_2$) are then approximated using a second order multivariate Taylor Expansion, such that
\begin{align*}
	var(\hat{d}_1) & = \Bigg[1-\frac{a\left\{(1-\rmp)^2+k\left(1-2(1-\rmp)\rmp\right)\right\}}{(1-\rmp)^2} - 2d_1N_2\\
	&\qquad +N_1\Bigg\{\frac{6\rmp-4\rmp^2-3}{(1-\rmp)^2} + e^{-d_1}\left(2+d_1\right)\\
	&\qquad\qquad+e^{-d_2}\left(1+\frac{\rmp^2}{(1-\rmp)^2} - d_1\right)+2d_1\Bigg\}\Bigg]\Bigg/\\
	&\qquad \Bigg[\frac{\left(1-a+N_1(e^{-d_2}+e^{-d_1}-2)\right)\left\{\frac{ak-N_1(e^{-d_2}-1)}{(1-\rmp)^2}+\frac{ak-N_1(e^{-d_1}-1)}{\rmp^2}\right\}}{H^2} \\
	&\qquad +e^{-2(H+d_1)}\Big\{e^{H+d1}\left(N_1-N_2\right)-0.5N_1e^{2d_1}+0.5N_1e^H\Big\}^2\Bigg].
\end{align*}
The variance for $\hat{d}_2$ is a mirror image of the variance for $\hat{d}_1$, with $R_{1,0}$ replaced by $1-R_{1,0}$, $N_1$ replaced with $N_2$, and so on.

One method to compare the variances of $\hat{\lambda}_j$ and $\hat{d}_j$ when the number of observed failures is small is to examine what happens to the variance as the number of observed failures goes to zero (i.e. as the cumulative hazard $H$ goes to zero).  To do this, the Poisson parameters are rewritten such that $\lambda_j$ is a function of $H$and $\xi_j$, the ``height" of the hazard rate in bin $j$ divided by the cumulative hazard (i.e. $\lambda_j = H\times\xi_j$).  In the two bin model, $\xi_1 = \rmp,$ and $\xi_2 = 1-\rmp$, and in general, $\xi_j = \mathcal{F}_j$ (as seen in equation (\ref{eqn:Fj})), where $\mathcal{F}_j$ represents the functions of the $R_{m,p}$ values associated with the hazard rate in bin $j$.  

In examining the limit of the variances as the cumulative hazard goes to zero, both $\lim_{H \to 0}var(\hat{d}_1) = 0,$ and $\lim_{H \to 0} var(\hat{\lambda}_1) = 0$, so it is important to determine which goes to zero ``faster" (i.e. for the same values of $H$, which variance is smaller).  To determine which function is smaller for the same values of $H$, we take the limit of the derivative of the difference between the two functions as $H$ goes to zero.  If we let $g_1(H) = var(\hat{d}_1(H)) - var(\hat{\lambda}_1(H))$, and $g_2(H) = var(\hat{d}_2(H)) - var(\hat{\lambda}_2(H))$ then 
\begin{align*}
	& \lim_{H \to 0} \left(\frac{\partial}{\partial H}g_1(H)\right) = -\frac{\rmp}{N_1}, & \lim_{H \to 0} \left(\frac{\partial}{\partial H} g_2(H)\right) & = -\frac{1-\rmp}{N_2}.
\end{align*}	
Since the derivatives of the differences are negative, the variance of the constant hazard rate for the MRH model is smaller than the variance for the constant hazard rate of the Poisson GLM as the number of observed failures becomes small.  

\begin{figure}[htbp] 
	\centering
	\includegraphics[width=6in]{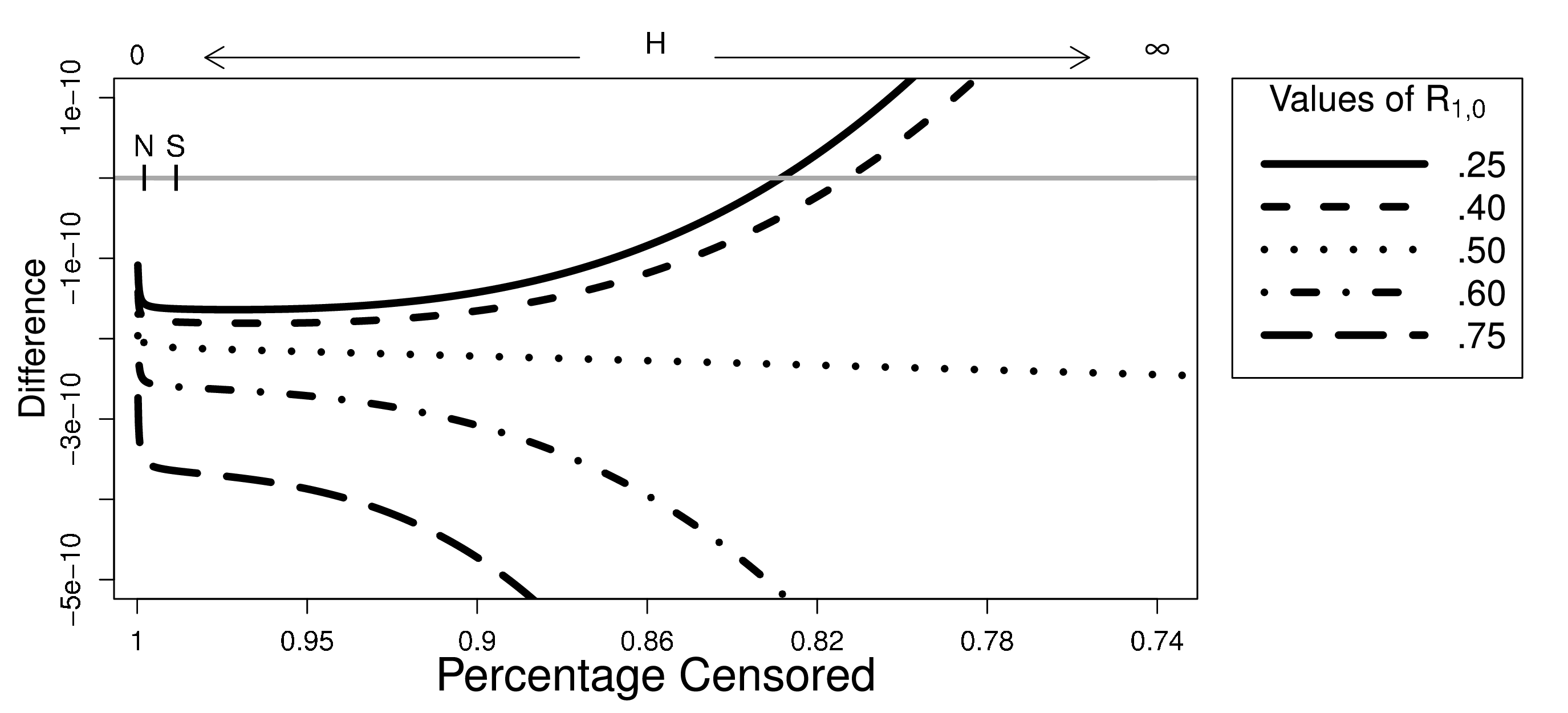}
	\caption{\footnotesize The difference of the variances for the MRH model and the Poisson GLM for the first of two bins as a function of $H$ (i.e. $g_1(H) = var(\hat{d}_1(H)) - var(\hat{\lambda}_1(H))$), with zero denoted by the solid grey line.  We see that if $H$ is small enough ($\approx H < 0.20$, censored $\approx 82\%$), then $g_1(H)$ is negative, regardless of the value of $\rmp$.  For larger values of $H$, the function is negative for larger values of $\rmp$.  This benefit is minimal in the 2-bin model, however, as a large $\rmp$ in the first bin requires a small $1-\rmp$ in the second bin.  The percentage of censored subjects in the simulated data sets (the average percent censored) is denoted with `S', and denoted with `N' for the Navrongo data set.  It can be observed that both percentages of censoring ($\approx 98\%$) are well above the threshold where the difference between the variances is always negative.}
	\label{fig:varDiffs}
\end{figure}

Figure \ref{fig:varDiffs} illustrates exactly how small $H$ needs to be for this to hold true by showing calculated values of $g_1(H)$ for censoring percentages ranging from 75\% to almost 100\% (corresponding to values of $H$ ranging from 0.30 to 0 using the relationship $S(t) = -\log H(t)$) and different values of $\rmp$.  It can be observed that for any value of $\rmp$, once $H$ is small enough (below about $H = 0.2$, or 82\% censored), the difference between the two variances is negative (with zero denoted by the solid grey line), meaning that $Var(\hat{d}_1) < Var(\hat{\lambda}_1)$.  When $H$ is greater than 0.2 (i.e. the censoring is less than about 82\%), the sign of the difference depends on the value of $\rmp.$  If $\rmp \ge 0.50$, the sign of the difference is negative for all values of $H$.  However, when $\rmp$ is greater than 0.50 (causing $g_1(H)$ to be negative), the next consecutive bin is associated with $1-\rmp$, which would be less than 0.5, in which case $Var(\hat{d}_2) > Var(\hat{\lambda}_2)$. With larger values of $H$, the two models will trade-off neighboring hazard estimates with the larger variances.  Denoted on the figure are the average percentage of censored subjects in the simulated data sets (`S') and the percentage of censored subjects in the Navrongo data set (`N'), which are both well below the threshold where the both $g_1(H)$ and $g_2(H)$ become negative.  In this regime, the MRH is more efficient with any prior choice for $\rmp$ when compared to the Poisson GLM.

\section{Estimator Assessment Through Simulation} 
To compare the accuracy and variance of the Poisson GLM, MRH, and the Poisson GAM estimators, we simulated data with weather and failure rate patterns similar to those in the Navrongo, Ghana data set, with about 0.019\% of subjects contracting meningitis yearly. The number of deaths in each bin was randomly sampled from a Poisson distribution with $\mu = 120$, and at the beginning of each time bin, the number of subjects at risk was 150,000 (the approximate average population of Navrongo over the 11-year period). The simulated data covered a 10-year period, with meningitis counts and weather variables measured at 16 equally spaced intervals each year.  The hazard rate used to generate the survival times had a peak mid-season (representing the peak season for meningitis).  In addition, we incorporated two time-varying weather covariates: the monthly average maximum daily temperature and the monthly average humidity at 3pm.  

We generated 200 data sets and examined the estimated results from all three models by comparing the mean, the 2.5\% and 97.5\% quantiles, the within-bin standard deviation, the bias and the RMSE of the 200 estimates.  The bias of the hazard rate was calculated for each model at each time point $t_j$ as 
\begin{eqnarray} bias(\hat{h}(t_j)) = 1/200 \sum_{i=1}^{200} (h(t_j)-\hat{h}(t_{ij})),\label{eqn:bias} \end{eqnarray}
and the RMSE of the hazard rate was calculated at each time point as 
\begin{eqnarray} RMSE(\hat{h}(t_j)) = \sqrt{1/200 \sum_{i=1}^{200} (h(t_j)-\hat{h}(t_{ij}))^2}.\label{eqn:rmse}\end{eqnarray}  
Each MRH model was run for 5000 MCMC iterations, with the first 500 iterations burned, and the remaining thinned by 10 to alleviate autocorrelation. Point estimates for the MRH model were calculated as the median of the marginal posterior distribution of each parameter.  The Poisson GAM is fitted via a penalized maximum likelihood with $g(t)$ estimated using penalized regression splines.  In the simulations, we use a basis dimension equal to 16 and specify a cyclic cubic regression spline to estimated the smooth underlying function, in order to match the value of the function at the end of the season (i.e. June) with the beginning of the season (i.e. July).

\begin{figure}[htbp] 
	\centering
	\includegraphics[width=5in]{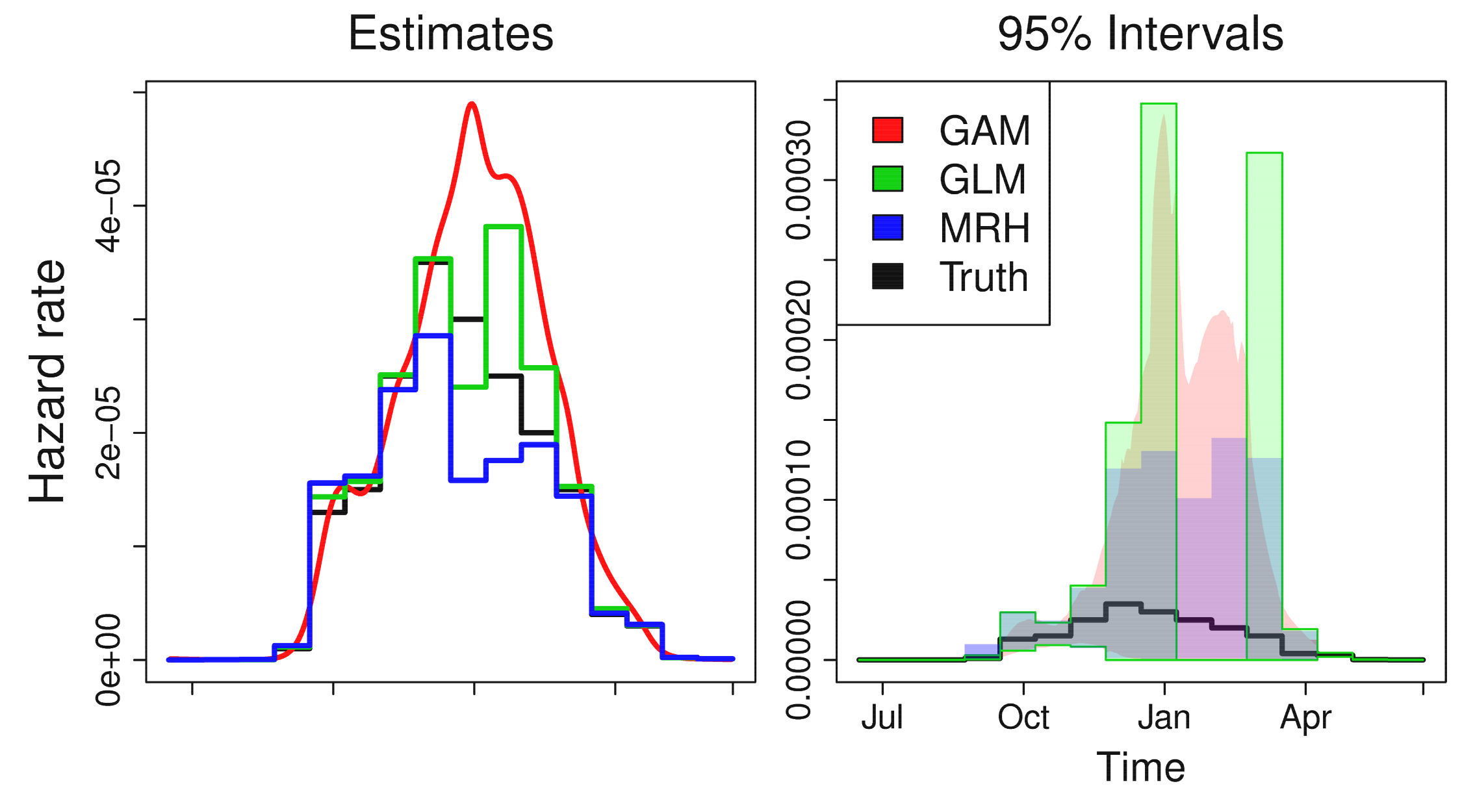}
	\includegraphics[width=5in]{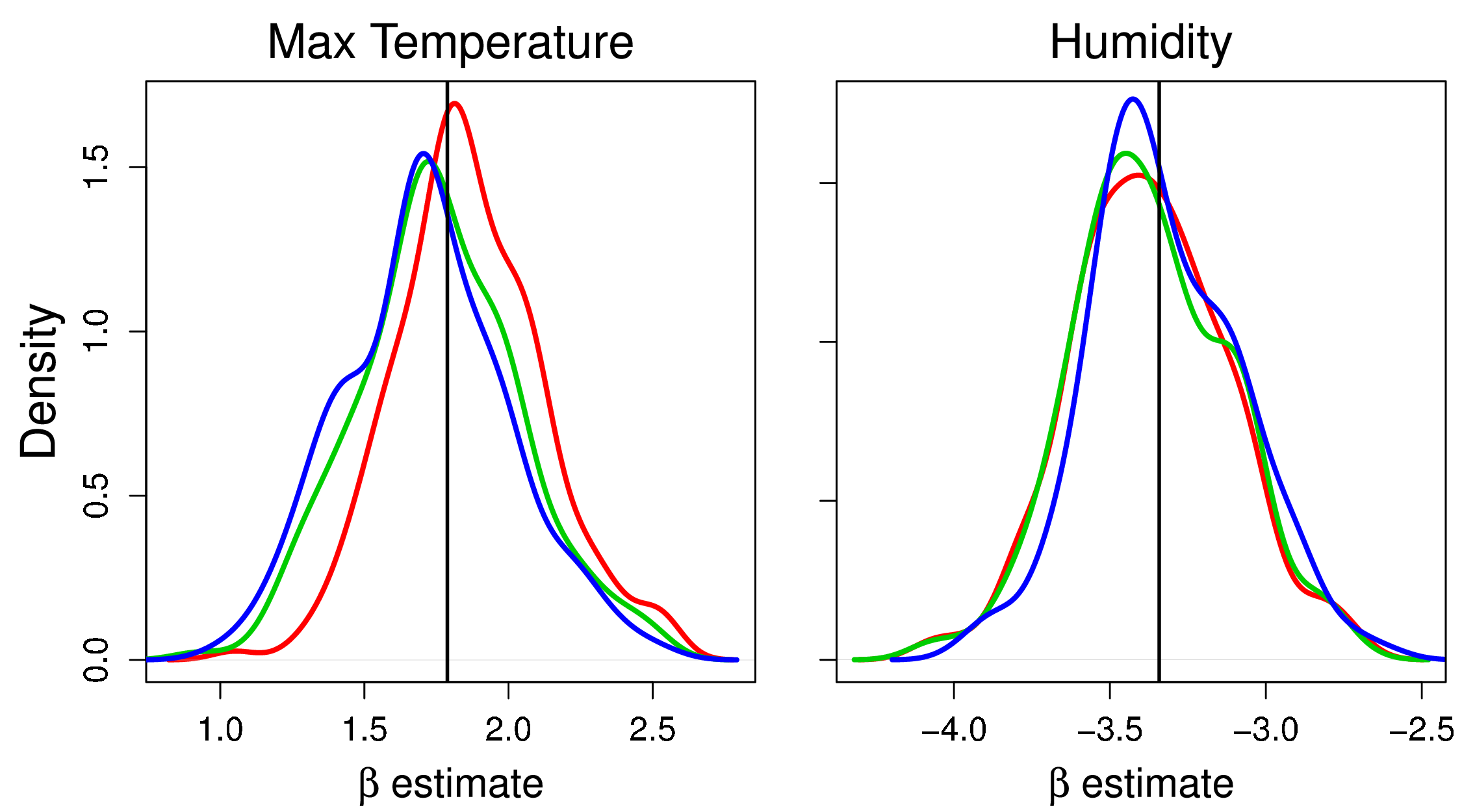}
	\caption{\footnotesize TOP:  The estimated hazard rates (left) from the three models, with the true hazard rate superimposed in black.  The mean of the 200 estimated hazard rates is shown for each model, and it can be observed that the Poisson MRH model tends to underestimate the true hazard rate, while the other two models tend to overestimate the true hazard rate.  On the left, 95\% intervals (calculated as the 2.5\% and 97.5\% quantiles of the 200 estimates) for all three models for the estimated hazard rates. The intervals are narrowest for the Poisson MRH models, and largest for the Poisson GLM.  This was expected based on the work shown in Section \ref{sec:varcompare}.  In addition, we see that while the mean of the Poisson GLM simulations is close to the true hazard rate, the quantiles are all quite small.  Both the Poisson GAM And MRH models have quantile bounds that capture the true hazard rate throughout the study. BOTTOM: Densities of the 200 estimated covariate effects for the average max daily temperature (left) and the average daily humidity at 3pm (right) for all three models, with the true parameter values superimposed as a black vertical line.  While the Poisson GAM models seem to deviate the most from the other two, there are few differences in the results of the three models.  The densities were calculated using the \texttt{density()} function in R. }
	\label{fig:simulEsts}
\end{figure}

The estimated hazard rates and covariate effects for all three models can be seen in Figure \ref{fig:simulEsts}.  The top left graph contrasts the mean of the 200 estimated hazard rates for each of the three models, with the true hazard rate shown in black.  All three models perform similarly, although the Poisson GAM seems to overestimate the hazard rate most frequently, and the MRH model seems to underestimate it most frequently.  The top right graph shows the 2.5\% and 97.5\% quantiles of the 200 estimates for each model.  The Poisson GLM bounds are the widest throughout most of the study period, with the exception between January and March, when the bounds are very narrow.  (Even though the mean of the 200 simulated data sets is close to the true hazard rate, the vast majority of the estimates are very small throughout this period of time, causing the 97.5\% quantile to be smaller than the mean.)  The 95\% intervals graph shows the MRH model has the smallest bounds.  Densities of the 200 estimated covariates effects (bottom graphs) show little differences among the three models. 

\begin{figure}[htbp] 
	\centering
	\includegraphics[width=6.5in]{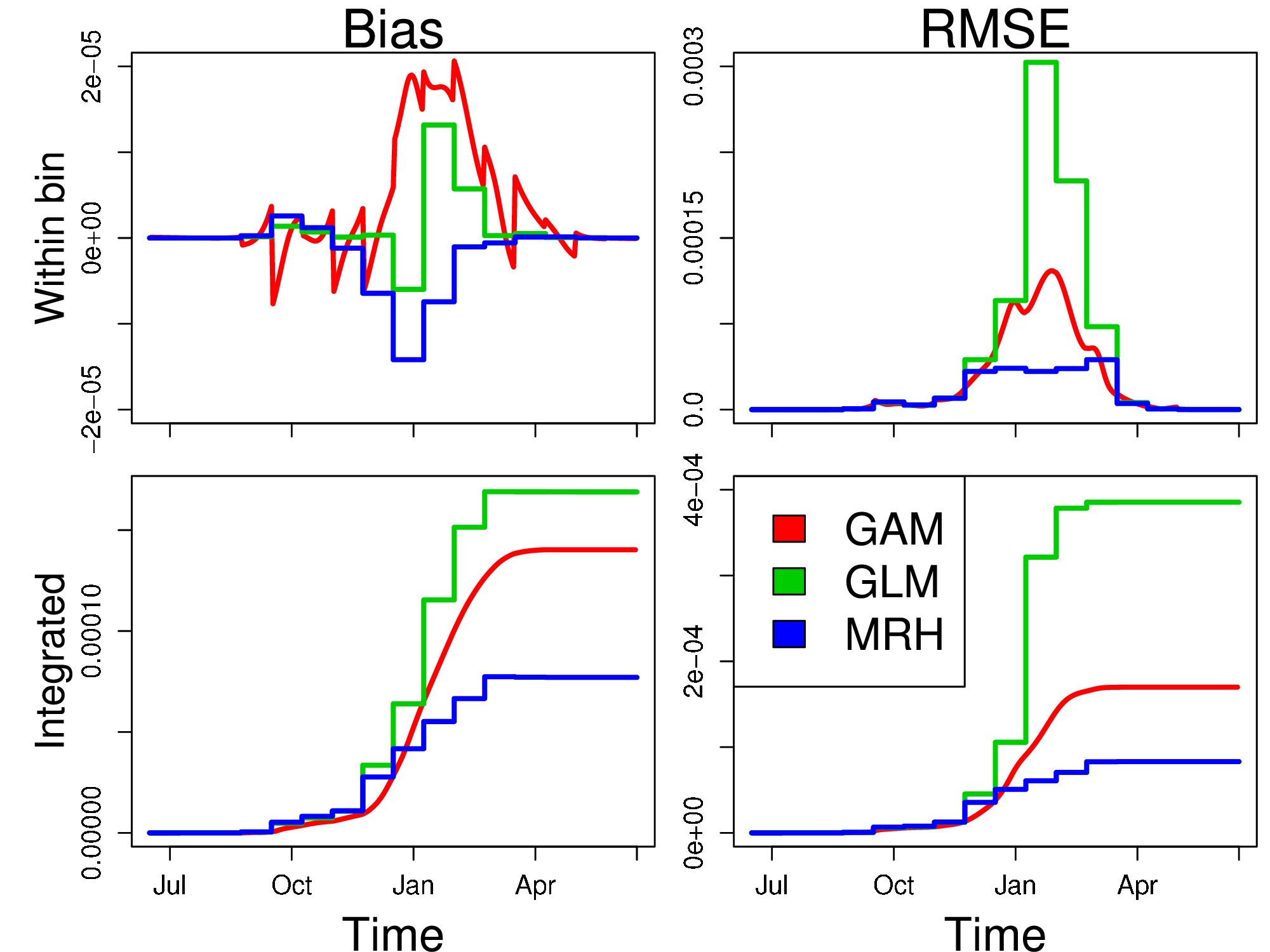}
	\caption{\footnotesize The bias and RMSE (calculated using equations (\ref{eqn:bias}) and (\ref{eqn:rmse})) over time for the hazard rate estimates from the 200 simulated data sets for the three models.  The top two graphs contain the bias and RMSE calculated within each bin, and show that the MRH model tends to underestimate the hazard rate, while the Poisson GAM and Poisson GLM tend to overestimate the hazard rate. The MRH model has a smaller absolute bias than the other two.  The RMSE calculations show that the MRH model has the smallest RMSE across the entire study period, followed by the Poisson GAM and Poisson GLM.  The integrated bias and RMSE (bottom graphs) tell a similar story, with the smallest integrated bias and RMSE values for the MRH model, he largest integrated bias for the Poisson GAM, and the largest integrated RMSE for the Poisson GLM.}
	\label{fig:simulbiasrmse}
\end{figure}

\begin{SCfigure}
	\centering
	\includegraphics[width=3.5in]{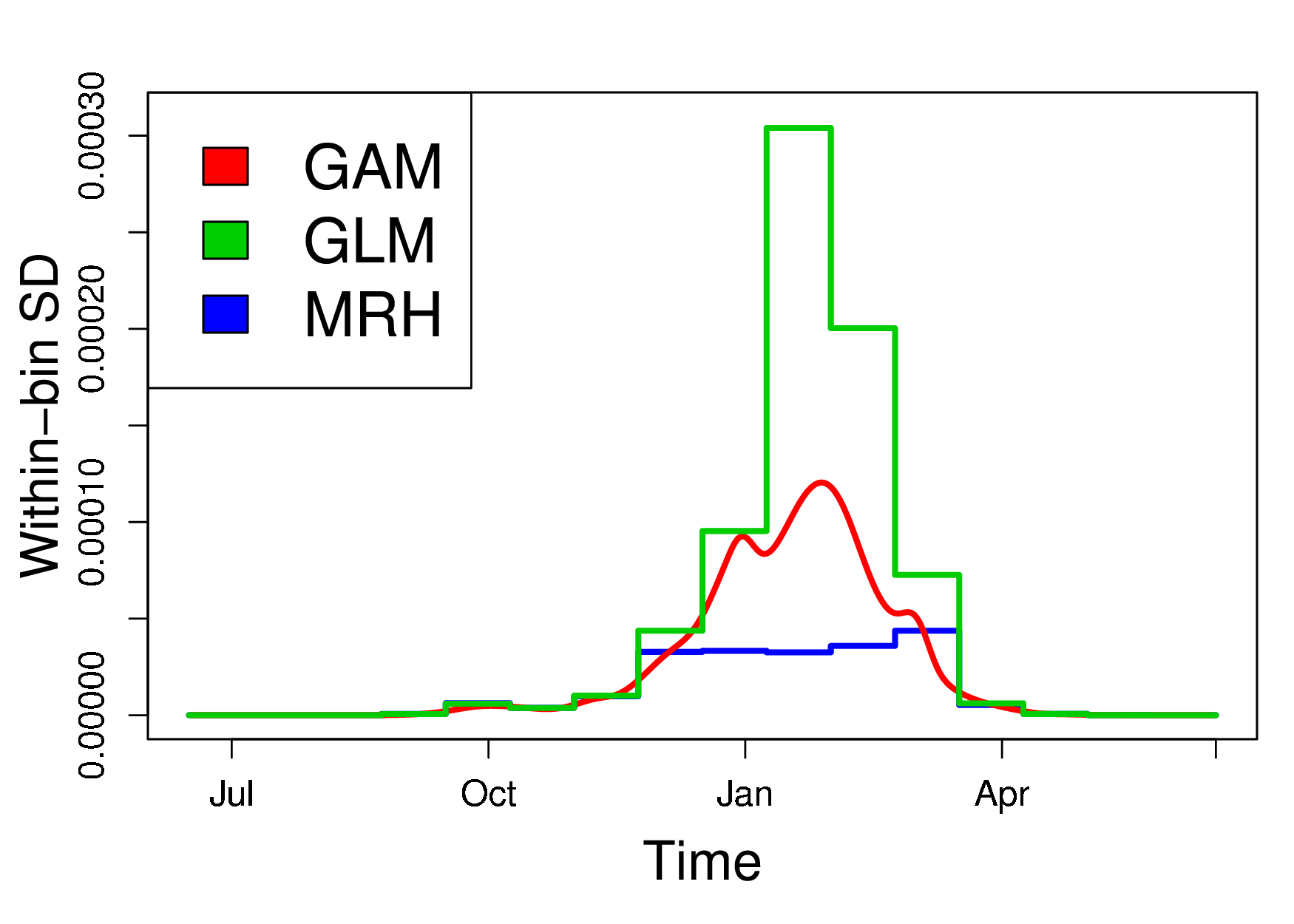}
	\caption{\footnotesize The within-bin standard deviation of each of the 200 estimates for each of the Poisson GLM, MRH model, and Poisson GAM.  As shown in Section \ref{sec:varcompare}, the MRH model has a smaller variance than the Poisson GLM.  From the simulations, it appears that the standard deviation of the estimator for the Poisson GAM is in between the other two.}
	\label{fig:SDs}
\end{SCfigure}

Figure \ref{fig:simulbiasrmse} shows the bias and RMSE within each bin (top graphs), and integrated over time (bottom graphs).  The within-bin bias shows that the MRH model tends to underestimate the hazard rate, while both the Poisson GLM and GAM tend to overestimate the hazard rate (top, left).  However, the MRH model has the smallest absolute bias, as well as the smallest RMSE, integrated absolute bias, and integrated RMSE across the study period.  Alternatively, the Poisson GLM has the largest RMSE, largest integrated absolute bias, and largest integrated RMSE.  Similarly, in Figure \ref{fig:SDs}, the within-bin standard deviation of the estimated hazard rates for each model are displayed.  As shown in Section \ref{sec:varcompare}, the MRH model has a smaller standard deviation when compared to the Poisson GLM.  It also appears that the Poisson GAM has a larger standard deviation than the MRH model, but a smaller standard deviation than the Poisson GLM.

\section{Analysis of Navrongo Data}
Navrongo, located in northern Ghana, experiences two main weather seasons, a wet season lasting from June to October, and then a dry season for the remainder of the year.  The population of Navrongo (located in the Kassena-Nankana district) has ranged from 140,000 to 150,000 residents, with approximately 20,000 people living in the city, and the remainder living in rural areas surrounding the city \cite{navrongopopn, oduro}.  The data set we examine contains monthly meningitis cases (laboratory confirmed) over an 11-year period from 1998-2008 collected by researchers at the Navrongo Health Research Centre (NHRC).  In addition to the meningitis case counts, our data set also contains meteorological data collected by Ghana Meteorological Services, and includes daily dust status, number of sunshine hours, maximum and minimum daily temperature, relative humidity, rain quantity, and wind speed. The data set also contains carbon monoxide (CO) emission estimates, calculated using the Fire INventory model (FINNv1; \cite{wiedinmyer}) based on satellite observations of active fires, and the number of confirmed pneumonia cases each month.  The weather data are given as monthly averages or percentages.  

While exact population counts were not available for each month, we estimated monthly counts as a linear increase over time, using population information provided in \cite{oduro, navrongopopn} to obtain the approximate number of deaths and births  (and hence the number of subjects at risk) each month.   Over the 11-year period, a total of 364 meningitis cases (range per year = 0 to 115, SD = 42.6) were documented, and the number of subjects at risk ranged from 141,046 in January, 1998 to 153,236 in December, 2008.  The set of covariates we examine in this manuscript was chosen based on previous analyses of this data set, which can be found in \cite{jabes}, where the authors consider multiple models, and discuss the best ones based on biological rationale and through model selection criteria (such as AIC \cite{refAIC}).  As such, our model includes the following predictors:
\begin{itemize}
	\item Maximum daily temperature (degree Celsius, monthly average).
	\item Minimum daily temperature in previous month (degree Celsius, monthly average).
	\item Percentage of days of the month with dust (daily measure is a ``yes" or ``no" binary variable).
	\item Average relative humidity at 3 pm in previous month (percent).
	\item Average CO emissions from fires in previous month (grams of CO per day). 
	\item Number of pneumonia cases two months prior. 
\end{itemize}
Certain variables (CO emissions, humidity, and minimum temperature) are thought to have a longer-term effect on the risk of meningitis, and are therefore lagged by one month, while the maximum daily temperature and the amount of dust in the environment are thought to directly impact current health.   The pneumonia cases are included as a proxy for respiratory health.  (See Table \ref{tab:summcateg} for a summary of the variables used in our analysis.)  
\begin{table}
	\caption{Sample characteristics of the 11 years of data gathered in Navrongo, Ghana, including summaries of the monthly weather averages and percentages and monthly meningitis counts.}\label{tab:summcateg}
\centering
\begin{tabular}{|l|cccc|}
	\hline
	Variable&Mean&Median&SD&Range\\
	\hline
	Monthly Meningitis Cases&2.8&0.0&8.3&(0.0, 42.0)\\
	Maximum Daily Temperature (Celsius)&35.4&36.0&3.1&(29.9, 41.1)\\
	Minimum Daily Temperature (Celsius)&23.0&22.9&2.2&(18.1, 28.3)\\
	Percent Relative Humidity (at 3pm)&39.5&36.5&20.6&(6.0, 73.0)\\
	Percentage dusty days&36.8&0.0&44.1&(0.0, 100.0)\\
	CO emissions (grams CO/day)&16.5&0.4&33.5&(0.0, 160.4)\\
	Monthly Pneumonia Cases&1.4&0.0&2.4&(0.0, 14.0)\\
	\hline
\end{tabular}
\end{table}

In addition to the weather variables, the Poisson GLM included indicators for each of the twelve months (to estimate the constant hazard in each bin), and was implemented using \texttt{glm()} with a poisson link in R.  The Poisson GAM was implemented using \texttt{gam()} in the \texttt{mgcv} library in R, with the smoothing performed using cyclic cubic regression splines (penalized cubic regression splines with ends that match the beginning and end of the season) and basis dimension equal to 12.  Offsets for both models are included as described in Sections \ref{sec:poissglm} and \ref{sec:poissgam}.

To facilitate the dyadic splitting of the MRH model, the yearly period was divided into 16 subintervals.  Since weather measures are only provided monthly, the measures were smoothed (using \texttt{smooth.spline()} in R) and then predicted (using \texttt{predict.smooth.spline()} in R) for the 16 mid-bin time points. Note that while this is a larger model compared to the other two models, there is little impact since the observations are aggregated monthly in the original data.  The MRH model ran for 1 million MCMC iterations, with the first 250,000 burned, and the remaining thinned by 100 to reduce autocorrelation.  The mean of the hyperpriors for $a$ and $b$ were fixed based on the Nelson-Aalen estimate of the cumulative hazard. The MRH model was implemented using an extension of the MRH package available in R\cite{MRHR}.  For all models, the ``year" was a seasonal year, and ranged from July to June so that meningitis outbreaks, which generally occur December to May were mid-season, avoiding smoothing issues near the boundary and enabling a continuous estimate of the function of seasonal meningitis trends.   

\begin{figure}[htbp] 
	\centering
	\includegraphics[width=6.5in]{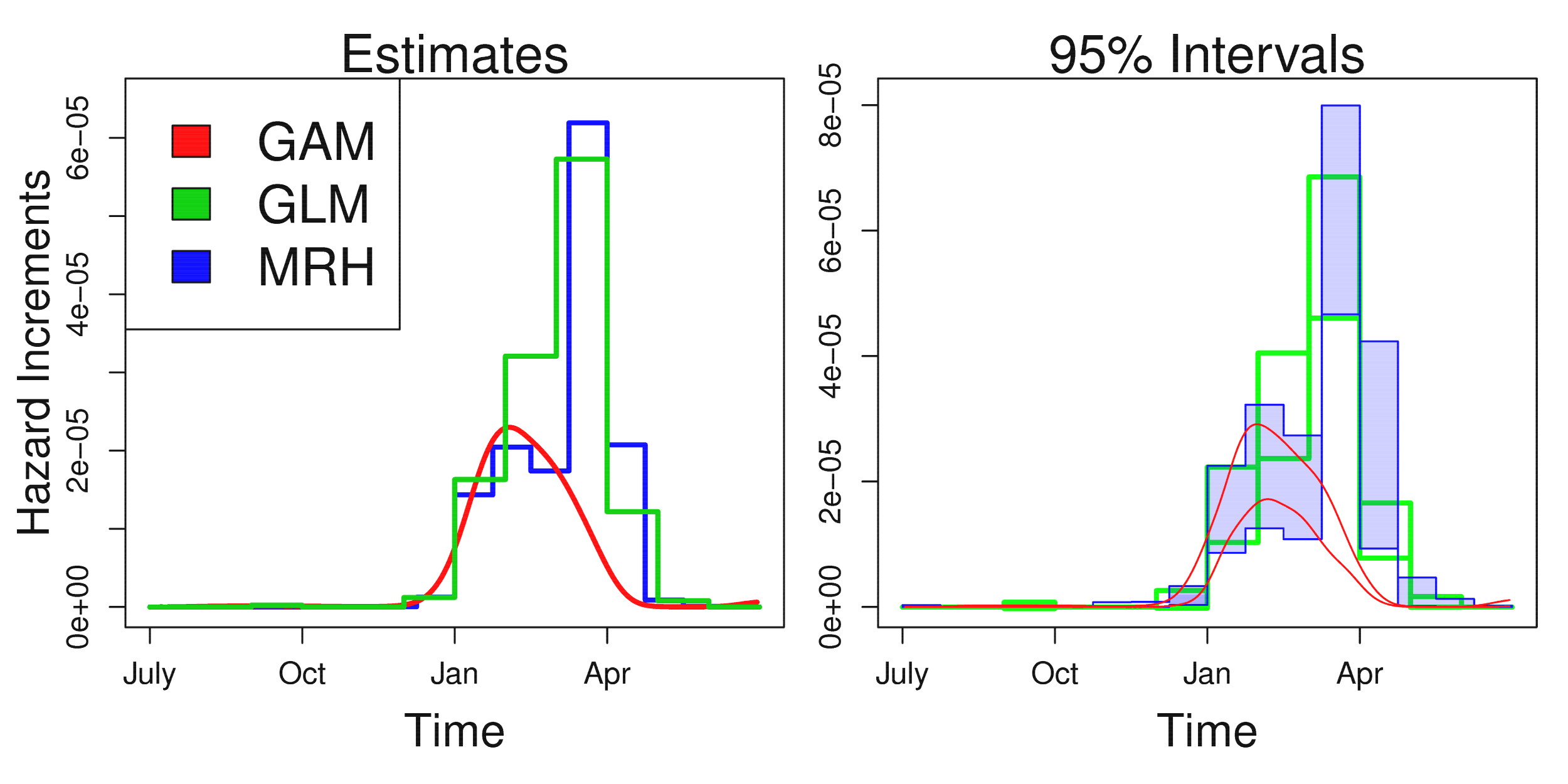}
	\includegraphics[width=6.5in]{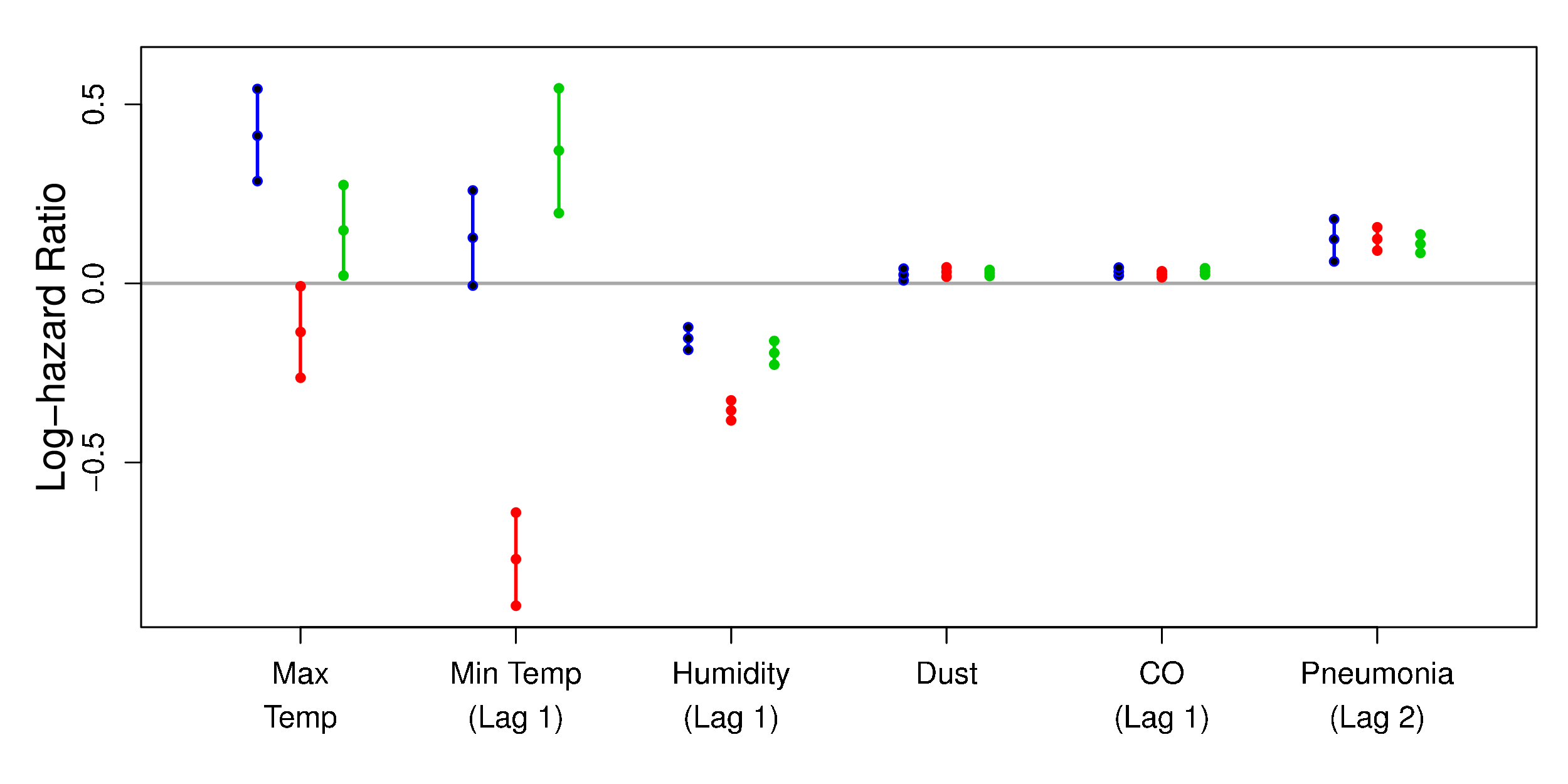}
	\caption{\footnotesize TOP: The estimated hazard rate of contracting meningitis (top left) and the 95\% interval bounds (top right, with credible intervals for the MRH model and confidence intervals for the GLM and GAM Poisson models).  While the MRH model hazard rate estimate differs slightly from the near-identical estimates of the Poisson GLM and GAM (left), the shape of all three models is similar: An increased risk of meningitis between January and April, with a peak in March, and virtually zero risk at other times of the season.  On the right, the 95\% interval bounds are shown.  While the Poisson GAM and MRH models have intervals with similar widths and shapes, the GLM model is dramatically (and unsurprisingly) larger across the entire peak of the season.  In addition, the bounds of the GLM go below zero, unlike the other two models.\newline
		BOTTOM:  The estimated effects of the weather covariates on the meningitis risk.  As expected, in all three models, increased daily temperatures (both minimum and maximum), percentage of dusty days, CO, and pneumonia cases increased the risk of meningitis.  Higher humidity, on the other hand, decreased the risk of meningitis.  All three models produced similar estimates and 95\% interval bounds (credible intervals for the MRH model, confidence intervals for the Poisson GLM and GAM), with the exception of the maximum and minimum daily temperatures.  The Poisson GLM and GAM are in agreement, with increases in minimum daily temperature having a larger affect on the increased meningitis risk than the maximum daily temperature.  However, the MRH model shows the opposite, with increases in the maximum daily temperature having a larger effect on the increased meningitis risk.  This is likely due to the fact that the two temperature covariates are somewhat collinear.}
	\label{fig:hazrateEst}
\end{figure}

The estimated hazard rate and covariate effects can be observed in Figure \ref{fig:hazrateEst}, which shows the average yearly hazard rate (top left), the 95\% error bounds (95\% credible intervals for the MRH model, and 95\% confidence intervals for the Poisson GLM and Poisson GAM, top right), and the estimated effects of the time-varying covariates (bottom).  The figure shows similar estimates for the hazard rate for the Poisson GLM and MRH model.  However, the Poisson GAM shows a much smaller hazard rate then the other two, although still displays the mid-season meningitis peak. The 95\% error bounds also seem similar among the models, although it is slightly difficult to compare the width of the GAM intervals with the other two models as it is smaller.  The bottom graph of the figure shows almost identical estimates and 95\% error bounds for all three models for the estimated effects of Dust, CO, and pneumonia, and minor differences with humidity.  However, the estimates for the effects of minimum and maximum daily temperature show some major differences.  While the Poisson GLM and MRH models show positive estimated effects for both temperature covariates, the MRH model shows a larger effect for the maximum daily temperature and a smaller effect for the minimum daily temperature, while the Poisson GLM shows the reverse.  The Poisson GAM shows negative estimated effects for both temperature covariates, which explains the differences in the estimate of the hazard rate. These discrepancies are likely due to the collinearity between the two temperature variables.

\begin{figure}[htbp] 
	\centering
	\includegraphics[width=6.5in]{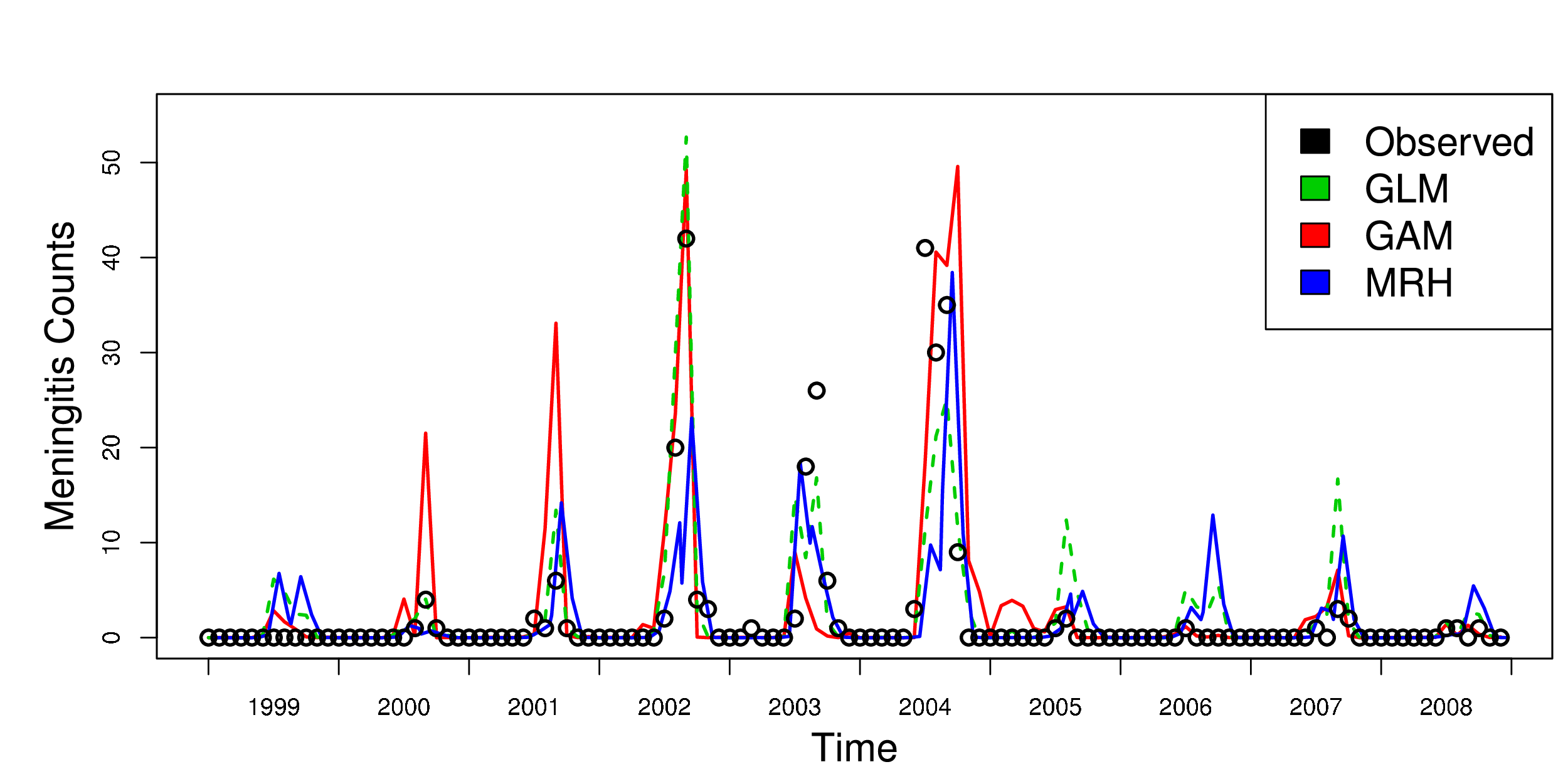}
	\caption{\footnotesize The estimated number of meningitis counts (i.e. the fitted values) over the 10-year seasonal period (July 1998 to July 2008), with the observed counts shown in black and the other models shown for comparison.  The Poisson GAM and GLM produce virtually identical estimates, while the MRH model differs slightly.  All three models tend to overestimate the small outbreaks (such as that in 1999, where there were zero meningitis confirmed meningitis cases, but all three models predicted at least a few).  The Poisson GLM and GAM accurately captured the 2002 outbreak, while the MRH model underestimated the number of cases.  However, in 2004, the MRH model accurately captured the high number of cases, while the other two models did not perform as well.}
	\label{fig:countsandbetas}
\end{figure}

\begin{figure}[htbp] 
	\centering
	\includegraphics[width=6.5in]{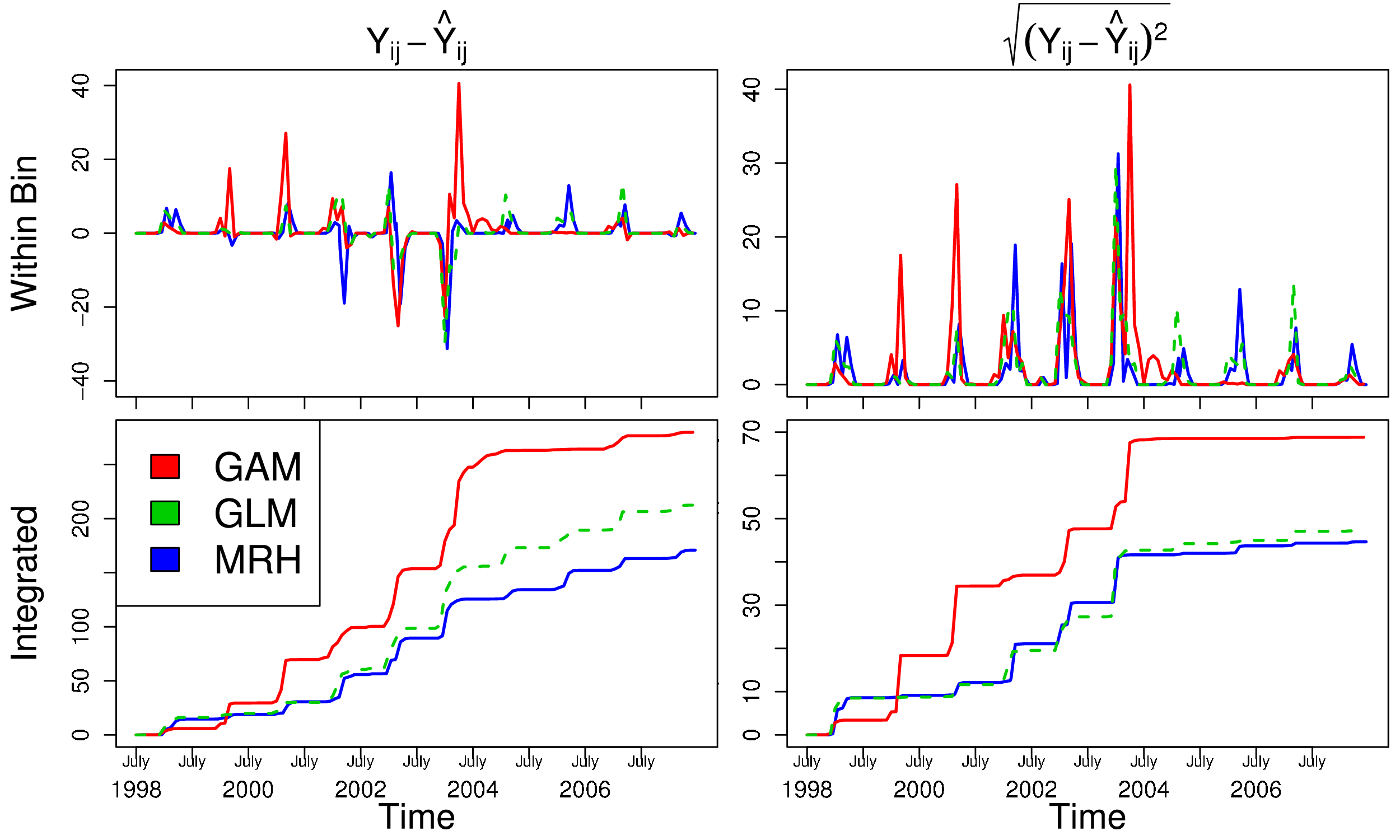}
	\caption{\footnotesize Differences between the observed and fitted values for the three models.  The within-bin values of the difference (top left), and the square root of the squared difference (top left) show only a little variation between the three models.  However, the integrated absolute difference (bottom left) and integrated squared difference (bottom right) show obvious, smaller integrated values for the Poisson MRH model.  The integrated values are the cumulative differences, multiplied by the bin width (equal to one for the Poisson GLM and GAM, and equal to 12/16 for the Poisson MRH).}
	\label{fig:residsreal}
\end{figure}

The fitted monthly meningitis counts, plotted against the observed monthly meningitis counts, are shown in Figure \ref{fig:countsandbetas}, where $\hat{Y}_{ij} =W_{ij}  \hat{\lambda}_{0j}\exp(X_i'\hat{\beta}+Z_{ij}'\hat{\alpha})$.  While all three models perform similarly, the Poisson GLM and MRH model seem to do a better job capturing the true peaks than the Poisson GAM, which seems to overestimate the number of cases when the outbreaks are larger (such as in years 2000 and 2001).  However, in years with very few meningitis counts, the Poisson GAM seems to be more reliable in capturing this than the other two models (such as in years 1999 and 2006). The difference between the observed and fitted values can be seen in Figure \ref{fig:residsreal}, which shows the monthly difference (top left), the square root of the monthly difference, squared (i.e. $\sqrt{\left(Y_{ij}-\hat{Y}_{ij}\right)^2}$, top right), the integrated absolute values of the monthly differences (bottom left), and the square root of the integrated square of the difference (bottom right).  The integrated values have been multiplied by their bin width.  It is hard to discern differences between the MRH model and Poisson GLM based on the within-bin residuals (top graphs), but it does seem more obvious that the Poisson GAM shows larger residuals than the other two models.  The cumulative graphs (bottom) show the smallest residual differences for the MRH model, and equal squared differences between the MRH model and Poisson GLM.  As shown in the top graphs, the Poisson GAM has the largest residuals of the three models.

\section{Discussion}
In this manuscript, we compare differences among the Poisson GLM, the MRH model, and the Poisson GAM.  While the models are all very similar in the likelihood formulation and set-up, we observed differences (some minor, some major) in their performance.  Because of its Bayesian structure and method of estimation, the MRH model provided the most accurate results in both the simulations and the analysis of the Navrongo data.  In addition, we showed both theoretically and through simulations that the MRH estimator has the smallest standard deviation.  However, the MRH model is more complex, and, as with most MCMC estimation routines, requires a longer run-time than the other two classical models.  The simulations show that the Poisson GAM, with a smooth function to estimate the hazard rate, performs better than the Poisson GLM.  However, in the real data analysis, the Poisson GLM had smaller residuals and did a better job predicting the meningitis counts over the 11-year period when compared to the Poisson GAM.

In the analysis of the Navrongo data, we are limited by the grouped data, and are unable to explore patterns in the contraction of meningitis at the subject-level.  Two important factors that cannot be accounted for are the location of the meningitis case (at the subject-level), as those living in the city likely have different risks than those living in the more rural parts of Navrongo, as well as seasonal migration patterns (which affect the estimated number at risk).  In addition, our analyses assume that the risk of contracting meningitis starts anew each season, and we cannot remove those who contracted meningitis from future cases.  While the effect of these assumptions is likely small, it is possible that some of the cases recorded are for the same individual and that these individuals are at higher risk than others.  Lastly, those who died from meningitis are also not recorded, and so may also be counted twice (once as a meningitis case, and then once as a censored subject), although this number is small and probably does not have much impact on the final results.

All models are sufficient in describing how weather and other predictors affect contraction of meningitis in the Sahel, and the choice of the best model should be based on the number of observed failures as well as time constraints for estimation.

\section*{Acknowledgements}
This work was supported by grants NSF-GEO 1211668, NSF-DEB 1316334, and NIH-R01GM096655.  The authors thank the researchers at NCAR for their helpful advice and input.

\section*{Appendix A: Full conditionals for the hyperparameters  $a, b, k$, and $\gamma_{m, p}$}
The parameters in the prior distributions of $H$ and all $R_{m,p}$s and $a, b, k$, and $\gamma_{m, p}$, can either be fixed at desired values, or treated as random variables with their own set of hyperpriors.   Below are the forms of these full conditional distributions for a specific set of hyperpriors we chose. The notation $\eta^-$ will be used to denote the set of all data and all parameters except for the parameter $\eta$ itself.  The full conditionals are as follows:
\begin{itemize}

\item If $a$ is given a zero-truncated Poisson prior, $\ds \frac{e^{-\mu_{a }}\mu_{a }^{a }}{a !\left(1-e^{-\mu_{a }}\right)}$ (chosen for computational convenience), the full conditional distribution for $a $ is:
\begin{equation*}
\begin{array}{l}
\ds \pi (a   \mid a ^{-}) \propto \ds \frac{H ^{a }\mu_{a }^{a }}{b^{a }(a -1)! a  !} \frac{\ds }{}
  \Pi_{m=1}^{M}\Pi_{p=0}^{2^{m-1}-1}\left\{\frac{ R_{ m,p }^{2\gamma_{ m,p} k ^m  a  }(1-R_{ m,p })^{2(1-\gamma_{ m,p} )k ^m  a  } }{\mrm{B}(2\gamma_{ m,p} k ^m  a  ,2(1-\gamma_{ m,p} )k ^m  a  )}\right\}
 \end{array}
 \label{eq:apost}
\end{equation*}

\item If the scale parameter $b$ in the gamma prior for the cumulative hazard function $H$ is given an exponential prior with mean $\mu_b$, the resulting full conditional is:
\begin{equation*}
\ds \pi(b  |b^{-})\propto \ds \frac{1}{b^{a}}\exp\left\{-\left(\frac{ H}{b}+\frac{b}{\mu_b}\right)\right\}
\end{equation*}

\item If  $k$ is given an exponential prior distribution with mean $\mu_{k}$,  the full conditional distribution for $k$ is  as follows:
\begin{equation*}
\begin{array}{l}
\ds \pi(k  \mid k^{-})\propto \Pi_{m=1}^{M}\Pi_{p=0}^{2^{m-1}-1}
 \ds \left\{\frac{ R_{m,p }^{2\gamma_{m,p}k^m  a }(1-R_{m,p })^{2(1-\gamma_{m,p})k^m  a } }{\mrm{B}(2\gamma_{m,p}k^m  a ,2(1-\gamma_{m,p})k^m  a )}\right\}
e^{-\frac{k}{\mu_{k}}}
\end{array}
\label{eq:kpost}
\end{equation*}

\item If a Beta($u$, $w$) prior is placed on each $\gamma_{m,p}$, the full conditional distribution for each $\gamma_{m,p}$ is proportional to:
\begin{equation*}
\ds
   \frac{ R_{m,p }^{2\gamma_{m,p} k^m  a }(1-R_{m,p })^{2(1-\gamma_{m,p})k^m  a } }{\mrm{B}(2\gamma_{m,p} k^m  a ,2(1-\gamma_{m,p})k^m  a )}
 \gamma_{m,p} ^{u -1}(1-\gamma_{m,p} )^{w -1} \\
\end{equation*}
\end{itemize}

\bibliographystyle{plain}

\end{document}